\documentclass[conference,10pt,letterpaper]{IEEEtran}

\ifCLASSINFOpdf
   \usepackage[pdftex]{graphicx}
\else
\fi

\usepackage{comment}

\usepackage{amsmath}
\usepackage{mathtools}

\usepackage{acronym}

\usepackage{amsfonts}
\usepackage{amsthm}
\usepackage{thmtools}

\usepackage{siunitx}
\usepackage{booktabs}

\usepackage{mleftright}
\usepackage{bm}
\usepackage{bbm}

\usepackage{tikz}
\usepackage{pgfplots}

\usepackage[textsize=tiny]{todonotes} %

\usepackage[font=footnotesize, skip=8pt]{caption}
\ifCLASSOPTIONcompsoc
  \usepackage[caption=false,font=normalsize,labelfont=sf,textfont=sf]{subfig}
\else
  \usepackage[caption=false,font=footnotesize]{subfig}
\fi

\newcommand\blfootnote[1]{%
  \begingroup
  \renewcommand\thefootnote{}\footnote{#1}%
  \addtocounter{footnote}{-1}%
  \endgroup
}

\usepackage{graphicx}
\usepackage{color}
\usepackage{pgf, tikz, pgfplots}
\usetikzlibrary{shapes,arrows.meta, calc, matrix, patterns}
\usepgfplotslibrary{groupplots}

\DeclareMathOperator*{\argmin}{\arg\!\min}
\DeclareMathOperator*{\argmax}{\arg\!\max}

\newcommand\cs[1]{{\tilde{#1}}}

\definecolor{KITgreen}{rgb}{0,.59,.51}
\definecolor{KITpalegreen}{RGB}{130,190,60} 
\definecolor{KITblack}{rgb}{0,0,0}
\definecolor{KITblue}{rgb}{.27,.39,.66}
\definecolor{KITred}{rgb}{.63,.13,.13}
\definecolor{KITpurple}{rgb}{.64,.06,.48}
\definecolor{KITcyan}{rgb}{.14,.63,.87}
\definecolor{KITyellow}{rgb}{.98,.89,0}
\definecolor{KITorange}{rgb}{.87,.60,.10}

\definecolor{EPcolor}{rgb}{.64,.06,.48}
\definecolor{EPiCGcolor}{RGB}{130,190,60} 
\definecolor{EPAcolor}{rgb}{.14,.63,.87}

\definecolor{CGcolor}{rgb}{0,.59,.51}
\definecolor{pCGcolor}{RGB}{130,190,60}
\definecolor{GScolor}{rgb}{.87,.60,.10}
\definecolor{GaBPcolor}{rgb}{.27,.39,.66}

    \acrodef{AWGN}[AWGN]{additive white Gaussian noise}
    \acrodef{BER}[BER]{bit error rate}
    \acrodef{BP}[BP]{belief propagation} %
    \acrodef{BiCG}[BiCG]{biconjugate gradient}
    \acrodef{CG}[CG]{conjugated gradient}
    \acrodef{CIT}[CIT]{condition number}
    \acrodef{CSI}[CSI]{channel state information}
    \acrodef{EP}[EP]{expectation propagation}
    \acrodef{FOM}[FOM]{full orthogonalization method}
    \acrodef{GaBP}[GaBP]{Gaussian belief propagation}
    \acrodef{GS}[GS]{Gauss-Seidel}
    \acrodef{KL}[KL]{Kullback-Leibler}
    \acrodef{LMMSE}[LMMSE]{linear minimum mean squared error}
    \acrodef{MAP}[MAP]{maximum a posteriori}
    \acrodef{MIMO}[MIMO]{multiple-input multiple-output}
    \acrodef{ML}[ML]{maximum likelihood}
    \acrodef{MP}[MP]{message passing}
    \acrodef{MUI}[MUI]{multi-user interference}
    \acrodef{NLOS}[NLOS]{non-line-of-sight}
    \acrodef{PAM}[PAM]{pulse amplitude modulation}
    \acrodef{pCG}[pCG]{preconditioned CG}
    \acrodef{QAM}[QAM]{quadrature amplitude modulation}
    \acrodef{RRE}[RRE]{relative residual error}
    \acrodef{SER}[SER]{symbol error rate}
    \acrodef{SNR}[SNR]{signal-to-noise ratio}
    \acrodef{UMa}[UMa]{urban macrocellular}

\pgfplotsset{compat=newest}
\begin{document}
\title{Fast and Robust Expectation Propagation MIMO Detection via Preconditioned Conjugated Gradient} %

\author{\IEEEauthorblockN{Luca Schmid\IEEEauthorrefmark{1}, Dominik Sulz\IEEEauthorrefmark{2}, and Laurent Schmalen\IEEEauthorrefmark{1}}
\IEEEauthorblockA{\IEEEauthorrefmark{1}Communications Engineering Lab, Karlsruhe Institute of Technology, 76187 Karlsruhe, Germany\\
\IEEEauthorrefmark{2}Mathematical Institute, University of Tübingen, 72076 Tübingen, Germany\\
Email: \IEEEauthorrefmark{1}\texttt{first.last@kit.edu}, \IEEEauthorrefmark{2}\texttt{dominik.sulz@uni-tuebingen.de}}
}

\markboth{Submitted version, \today}%
{Schmid \MakeLowercase{\textit{et al.}}: XXX}

\maketitle
\begin{abstract}
We study the expectation propagation (EP) algorithm for symbol detection in massive multiple-input multiple-output (MIMO) systems. The EP detector shows excellent performance but suffers from a high computational complexity due to the matrix inversion, required in each EP iteration to perform marginal inference on a Gaussian system. We propose an inversion-free variant of the EP algorithm by treating inference on the mean and variance as two separate and simpler subtasks: 
We study the preconditioned conjugate gradient algorithm for obtaining the mean, which can significantly reduce the complexity and increase stability by relying on the Jacobi preconditioner that proves to fit the EP characteristics very well.
For the variance, we use a simple approximation based on linear regression of the Gram channel matrix.
Numerical studies on the Rayleigh-fading channel and on a realistic 3GPP channel model reveal the efficiency of the proposed scheme, which offers an attractive performance-complexity tradeoff and even outperforms the original EP detector in high multi-user inference cases where the matrix inversion becomes numerically unstable.
\blfootnote{This work has received funding in part from the European Research Council (ERC) under the European Union’s Horizon 2020 research and innovation programme (grant agreement No. 101001899) and in part from the German Federal Ministry of Education and Research (BMBF) within the project Open6GHub (grant agreement 16KISK010).}
\end{abstract}

\IEEEpeerreviewmaketitle
\section{Introduction}
Massive \ac{MIMO} is discussed as a key technology for future cellular communication systems due to its ability to serve multiple users on the same time and frequency resource with a high spectral efficiency by leveraging diversity and multiplexing gains whose advantages scale with the number of antennas~\cite{yang_fifty_2015}.
Due to strong \ac{MUI}, efficient symbol detection becomes a crucial bottleneck in achieving the promising theoretical guarantees in practice. Optimal detection based on the \ac{ML} criterion quickly becomes intractable for growing system sizes. On the other hand, linear schemes with reduced complexity, like the \ac{LMMSE} detector, show poor performance.

Many proposed \ac{MIMO} detectors with a promising performance-complexity tradeoff are based on iterative algorithms that can be described as \ac{MP} on probabilistic graphical models~\cite{wu_low-complexity_2014}. 
A high-accuracy \ac{MP} detector is based on the \ac{EP} algorithm~\cite{minka_expectation_2013, cespedes_expectation_2014}. To obtain global consistency between local approximations, the \ac{EP} detector iteratively constructs a Gaussian approximation to the posterior distribution of the transmitted \ac{MIMO} symbols to match the moments in each \ac{MIMO} dimension.
It thereby achieves excellent performance for various antenna configurations~\cite{cespedes_expectation_2014}.

A major drawback of the \ac{EP} detector is the cubic complexity of the required matrix inversion in each \ac{EP} iteration, which becomes a prohibitive computational burden for future massive \ac{MIMO} systems with 32-256 antennas.
To avoid the expensive computation of the matrix inverse, the authors in~\cite{tan_low-complexity_2019} reformulate the \ac{EP} update equations using a fixed point approximation. 
The proposed approximation is quite coarse and the resulting matrix inversion-free detector suffers from slow convergence rate and severe performance degradation~\cite{tan_low-complexity_2019}. Therefore, the authors suggest falling back to the matrix inverse only for the initialization of the algorithm. The improved detector EPANet is proposed in~\cite{ge_improving_2021}, where deep learning techniques are considered to optimize the detector in an end-to-end manner. The EPANet algorithm works very well for low-\ac{MUI} scenarios but the performance gap to the \ac{EP} detector increases in the presence of high \ac{MUI}.
On the other hand, various works in literature study methods for approximating the matrix inverse, e.g., based on the Neumann-series~\cite{zhang_expectation_2018}. However, the complexity reduction is limited if a good approximation is intended. 

In contrast to the works mentioned above, we argue that the \ac{EP} detector does \emph{not} require the explicit computation of the full matrix inverse. Instead, the problem can be divided into the simpler subtasks of finding the diagonal elements of the inverse as well as solving a linear system of equations. 
The latter is a rigorously studied problem and there exists a variety of efficient approximative methods such as the \ac{GS} or the \ac{CG} algorithm~\cite{GvL13}. These methods were already considered for the \ac{MIMO} detection problem to approximate the \ac{LMMSE} detector~\cite{dai_low-complexity_2015, wei_learned_cg_2020} and in~\cite{takeuchi_rigorous_2017}, the error of the \ac{CG} algorithm was studied theoretically in the context of \ac{EP} for the asymptotic case.

In this work, we compare the suitability of different approximative linear solvers in relieving the computational burden in the \ac{EP} detector.
Leveraging an intrinsic property of \ac{EP}, we propose a well-suited preconditioner to substantially reduce the required number of \ac{CG} iterations.
For the diagonal elements of the inverse, we rely on a simple, yet effective approximation based on linear regression of the Gram channel matrix.
Numerical studies with varying \ac{MUI} scenarios on the Rayleigh fading channel as well as on a more realistic 3GPP 3D MIMO \ac{UMa} channel model demonstrate the effectiveness of the novel method and provide valuable insights into both applied approximations.
The proposed algorithm proves to be numerically more stable while achieving competitive results to the original \ac{EP} detector with lower complexity.

\section{System Model and Preliminaries}
\subsection{System Model} \label{subsec:channel_model} %
We consider the multi-user \ac{MIMO} channel, where $N_\text{t}$ single-antenna users simultaneously transmit information to a base station that is equipped with $N_\text{r}$ antennas. The observation~${\cs{\bm{y}} \in \mathbb{C}^{N_\text{r}}}$ at the receiver is modeled by
\begin{equation*}
    \cs{\bm{y}} = \cs{\bm{H}} \cs{\bm{x}} + \cs{\bm{n}},
\end{equation*}
where ${\cs{\bm{H}} \in \mathbb{C}^{N_\text{r} \times N_\text{t}}}$ is the channel matrix, ${\cs{\bm{x}} \in \cs{\mathcal{M}}^{N_\text{t}}}$ is the vector of transmitted symbols and ${\cs{\bm{n}} \sim \mathcal{CN}(\cs{\bm{n}}:\bm{0},2\sigma^2 \bm{I}_{N_\text{r}})}$ is an additive circular-symmetric complex Gaussian noise vector. The transmit symbols are independently and uniformly sampled from a \ac{QAM} constellation~${\cs{\mathcal{M}} \subset \mathbb{C}}$ of size ${|\cs{\mathcal{M}}|=M}$. We define the \ac{SNR} in dB at the receiver as
\begin{equation*}
    \mathsf{snr} := 10 \log_{10} \mleft( \frac{\mathbb{E} \big[ \lVert \cs{\bm{H}} \cs{\bm{x}} \rVert_2^2 \big]} {\mathbb{E} \big[ \lVert \cs{\bm{n}} \rVert_2^2 \big]} \mright).
\end{equation*}
The complex-valued system can be decomposed into an equivalent real-valued representation
\begin{equation*}
    {\bm{y}} = {\bm{H}} {\bm{x}} + {\bm{n}}, \quad \bm{H} = 
     \begin{pmatrix} \text{Re}\{\cs{\bm{H}}\} & - \text{Im}\{\cs{\bm{H}}\} \\ \text{Im}\{\cs{\bm{H}}\} & \text{Re}\{\cs{\bm{H}}\} \end{pmatrix}\in \mathbb{R}^{2N_\text{r} \times 2N_\text{t}}
\end{equation*}
with ${\bm{y} = (\text{Re}\{\cs{\bm{y}}\}, \text{Im}\{\cs{\bm{y}}\})^{\rm T}}$, ${\bm{x} = (\text{Re}\{\cs{\bm{x}}\}, \text{Im}\{\cs{\bm{x}}\})^{\rm T}}$ and ${\bm{n} = (\text{Re}\{\cs{\bm{n}}\}, \text{Im}\{\cs{\bm{n}}\})^{\rm T}}$, which will become convenient later. The complex-valued \ac{QAM} constellation $\cs{\mathcal{M}}$ is accordingly transformed to a real-valued \ac{PAM} constellation~$\mathcal{M}$.

The goal of \ac{MIMO} detection is to infer the transmit symbols~$\bm{x}$ from the observation~$\bm{y}$ at the receiver.
The posterior distribution can be expressed as
\begin{equation}
    p(\bm{x}|\bm{y}) \propto p(\bm{y}|\bm{x}) \prod\limits_{n=1}^{2 N_\text{t}} \mathbbm{1}_{\{x_n \in \mathcal{M}\}}, \qquad \bm{x} \in \mathbb{R}^{2 N_\text{t}} \label{eq:map_factorized} %
\end{equation}
where the indicator functions~${\mathbbm{1}_{\{x_n \in \mathcal{M}\}}}$ constrain the real-valued variables~$x_n$ to takes discrete values of the constellation~$\mathcal{M}$.
Optimal detection with respect to the \ac{SER} is based on the symbol-wise \ac{MAP} estimate, i.e., on the marginal distributions of eq.~\eqref{eq:map_factorized}:
\begin{equation}
    \hat{x}_{n,\text{ML}}  = \argmax\limits_{x_n \in \mathbb{R}} p(\bm{y} | x_n) \, \mathbbm{1}_{\{x_n \in \mathcal{M}\}}. \label{eq:map}
\end{equation}
We assume the availability of \ac{CSI}, i.e., the receiver has perfect knowledge of $\bm{H}$ and $\sigma^2$.
For the modeling of $\bm{H}$, we follow two approaches:

\subsubsection{Rayleigh-fading Channel Model}
Each element of $\cs{\bm{H}}$ is independently sampled from a circular complex standard Gaussian distribution $\mathcal{CN}(0,1)$. Albeit the rather unrealistic assumption of uncorrelated channel coefficients in practice, this channel model is well-studied, enabling a good comparison of this work to other methods in the literature, like~\cite{cespedes_expectation_2014, tan_low-complexity_2019, khani_adaptive_2020}.

\subsubsection{3GPP 3D MIMO UMa NLOS Channel Model}
For a more realistic model, we consider an \ac{UMa} scenario from the 3GPP 3D \ac{MIMO} channel model~\cite{3gpp:channel_model} with a \ac{NLOS} propagation.
We define a configuration similar to~\cite{khani_adaptive_2020}: $N_\text{t}$ single-antenna users are randomly dropped within a radius of $10$-$500\,$m from a base station in a 120°-cell sector and follow a linear trajectory with a constant speed of $30\,$km/h ($8.3\,$m/s). The base station is equipped with $N_\text{r}/2$ dual-polarized antennas that are installed in a rectangular planar array at $25\,$m height. The center frequency is set to $2.53\,$GHz, and the bandwidth of $20\,$MHz is divided among 1024 sub-carriers.
As opposed to~\cite{khani_adaptive_2020}, this work does not focus on the exploitation of correlations in the frequency and time domain. To sample more independent channel realizations, we subsample the sub-carriers with a rate of $32$ and only sample 1 OFDM symbol per channel realization.
Furthermore, we assume perfect power control, i.e., 
\begin{equation*}
    \frac{1}{N_\text{r}} \sum\limits_{i=1}^{N_\text{r}} \lVert \cs{\bm{H}}_{ij} \rVert_2^2 = 1, \quad j=1,\dots,N_t.
\end{equation*}

\subsection{Expectation Propagation for MIMO Detection}\label{subsec:EPdetector}
Exact \ac{ML} detection in~\eqref{eq:map} is typically intractable because marginalization on ${p(\bm{x}|\bm{y})}$ becomes prohibitive for ${N_\text{t} \gg 1}$.
\Acf{EP} is an approximate inference method that simplifies exact Bayesian inference by approximating posterior distributions with members of an exponential family~\cite{minka_expectation_2013}. 
In the remainder of this section, we discuss the application of the \ac{EP} method for \ac{MIMO} detection as it was proposed by Céspedes et al.\ in \cite{cespedes_expectation_2014}. 
For an excellent in-depth treatment of the general \ac{EP} method, we refer the reader to~\cite{bishop_pattern_2006} and~\cite{Seeger2005ExpectationPF}.

To simplify the distribution in \eqref{eq:map_factorized}, we approximate $\mathbbm{1}_{\{x_n \in \mathcal{M}\}}$ by a 1-dimensional Gaussian in canonical form
\begin{equation*}
    \mathbbm{1}_{\{x_n \in \mathcal{M}\}} \approx \exp \mleft( \gamma_n x_n - \frac12 \Lambda_n x_n^2 \mright) =: t_n(x_n),
\end{equation*}
with free parameters~$\gamma_n$ and ${\Lambda_n>0}$.
This significantly facilitates inference, since the resulting distribution
\begin{equation*}
    q(\bm{x}) := \mathcal{N}(\bm{y}: \bm{Hx}, \sigma^2\bm{I}_{2N_\text{r}}) \prod\limits_{n=1}^{2N_\text{t}} t_n(x_n)
\end{equation*}
belongs to the Gaussian exponential family~$\mathcal{F}_\text{G}$ for which inference is tractable~\cite{bishop_pattern_2006}.
In this context, the meaning of tractable inference is twofold: First, we can compute the moment parameters associated with the sufficient statistics of $q(\bm{x})$, i.e., the covariance matrix and mean
\begin{align}
    \bm{\Sigma} &= \left( \sigma^{-2} \bm{H}^{\rm T}\bm{H} + \text{diag}(\bm{\Lambda}) \right)^{-1}, \label{eq:matrix_inverse} \\ 
    \bm{\mu} &= \bm{\Sigma} \left( \sigma^{-2} \bm{H}^{\rm T} \bm{y} + \bm{\gamma} \right), \label{eq:ep_mean_computation_with_inverse}
\end{align}
where we have introduced the vector notations ${\bm{\Lambda}=(\Lambda_1,\ldots,\Lambda_{2N_\text{t}})^{\rm T}}$ and ${\bm{\gamma}=(\gamma,\ldots,\gamma_{2N_\text{t}})^{\rm T}}$. 
Second, \emph{marginal} inference for~$\mathcal{F}_\text{G}$ is tractable and is directly given by
${q(x_n) = \mathcal{N}(x_n: \mu_n, \Sigma_{nn})}$.
\ac{EP} is an iterative method to determine the parameters $\gamma_n$ and ${\Lambda_n > 0}$, ${n=1,\ldots,2N_\text{t}}$ such that the \ac{KL} divergence $D_\text{KL}\mleft(p(\bm{x}|\bm{y}) \Vert q(\bm{x}) \mright)$ is minimized. As shown in~\cite{bishop_pattern_2006}, this is equivalent to matching the expected sufficient statistics of ${p(\bm{x}|\bm{y})}$ and $q(\bm{x})$, also known as \emph{moment matching}. 
The \ac{EP} algorithm approaches the moment matching condition by sequentially matching the moments of marginal distributions in the factorization in~\eqref{eq:map_factorized}.
In each iteration~$\ell$, the parameter updates ${\gamma_n^{(\ell+1)}}$ and ${\Lambda_n^{(\ell+1)}}$ are chosen such that the moments of the marginal Gaussian distribution
\begin{equation*}
    q^{(\ell)}(x_n) \, \frac{ \exp \mleft( \gamma_n^{(\ell+1)} x_n - \frac12 \Lambda_n^{(\ell+1)} x_n^2 \mright)}{ \exp \mleft( \gamma_n^{(\ell)} x_n - \frac12 \Lambda_n^{(\ell)} x_n^2 \mright)} 
\end{equation*}
match with the mean~$\mu_{p_n}^{(\ell)}$ and variance~$\sigma_{p_n}^{2(\ell)}$ of the marginal tiled distribution
\begin{equation*}
    q^{(\ell)}(x_n) \, \frac{\mathbbm{1}_{\{x_n \in \mathcal{M}\}}}{ \exp \mleft( \gamma_n^{(\ell)} x_n - \frac12 \Lambda_n^{(\ell)} x_n^2 \mright)},
\end{equation*}
for all ${n=1,\ldots,2N_\text{t}}$, respectively.
We denote by ${q^{(\ell)}(x_n)}={\mathcal{N}(x_n:\mu_n^{(\ell)},\sigma_n^{2(\ell)})}$ the marginals of $q^{(\ell)}(\bm{x})$ in iteration~$\ell$.
Solving the moment matching conditions for ${\gamma_n^{(\ell+1)}}$ and ${\Lambda^{(\ell+1)}}$ leads to the \ac{EP} update equations
\begin{align}
    \Lambda_n^{(\ell+1)} &= \beta \left( \frac{1}{\sigma_{p_n}^{2(\ell)}} - \frac{1-\sigma_n^{2(\ell)} \Lambda_n^{(\ell)}}{\sigma_n^{2(\ell)}}\right) + (1-\beta) \Lambda_n^{(\ell)}, \label{eq:lambda_update} \\ 
    \gamma_n^{(\ell+1)} &= \beta \left( \frac{\mu_{p_n}^{(\ell)}}{\sigma_{p_n}^{2(\ell)}} - \frac{\mu_n^{(\ell)}}{\sigma_n^{2(\ell)}} + \gamma_n^{(\ell)}\right) + (1-\beta)\gamma_n^{(\ell)}, \nonumber
\end{align}
where ${\beta \in [0,1]}$ is a smoothing parameter.
After convergence or a fixed number of \ac{EP} iterations~$L$, the obtained approximation~${q(\bm{x}) \approx p(\bm{x}|\bm{y})}$ significantly simplifies the \ac{MIMO} detection problem in~\eqref{eq:map} to a nearest neighbor decision
\begin{equation*}
     \hat{x}_{n,\text{EP}} = \argmin\limits_{x_n \in \mathcal{M}} \big\lVert x_n - \mu_n^{(L)} \big\rVert_2^2.
\end{equation*} 

\section{The EPiCG Algorithm} \label{sec:EPiCG}
In this section we give a detailed description of the proposed \textit{Expectation Propagation integrated Conjugate Gradient} (EPiCG) algorithm.
To perform marginal inference on the multivariate Gaussian distribution~$q(\bm{x})$, i.e., to obtain the mean~$\mu_n^{(\ell)}$ and variance~$\sigma_n^{2(\ell)}$ of the marginals~$q^{(\ell)}(x_n)$, the \ac{EP} detector computes the covariance matrix~$\bm{\Sigma}$ in~\eqref{eq:matrix_inverse} for each iteration~$\ell$.
The matrix inversion operation is unfavorable from a numerical point of view since it can lead to numerical instabilities and further imposes a complexity of the order $\mathcal{O}((2N_\text{t})^3)$. This is typically unaffordable for modern massive \ac{MIMO} system sizes and thus inhibits widespread employment of the \ac{EP} detector in practice. In addition, the solution can become less accurate for ill-conditioned channel matrices, limiting the detection performance~\cite{kosasih_graph_2022}.

We argue that marginal inference on~$q(\bm{x})$ does \emph{not} require the computation of the \emph{full} matrix inverse. 
We only require the diagonal elements of the inverse to obtain the variance of the marginals~$q^{(\ell)}(x_n)$; their mean is provided as the solution of a linear system of equations.\footnote{Note that the \ac{GaBP} algorithm can provide the mean and variance of the marginals in one \ac{MP} algorithm and thus seems predestined for this problem. However, due to the underlying fully connected graph, we observed non-convergent behavior in most cases.}
Based on this observation, we propose to divide the  marginal inference problem into two separate and simpler sub-tasks.
In the following, we discuss both subtasks and propose approximative solutions with reduced complexity. Note that although we are treating both problems separately, the intrinsic moment matching condition of the \ac{EP} algorithm combines both tasks in each \ac{EP} iteration~$\ell$.

\subsection{Marginal Inference: Mean}\label{subsec:mean_approx}
As an alternative to~\eqref{eq:matrix_inverse}, the mean values~$\mu_n^{(\ell)}$ can be computed without the matrix inversion by solving the linear system of equations
\begin{align} %
    \bm{A}^{(\ell)}\bm{\mu}^{(\ell)} = \bm{b}^{(\ell)}, \quad \ell=0,\ldots,L \label{eq:linear_system}
\end{align}
with 
\begin{align}
    \bm{A}^{(\ell)} &:= \sigma^{-2} \bm{H}^{\rm T}\bm{H} + \text{diag}(\bm{\Lambda}^{(\ell)}), \label{eq:A_def} \\
    \bm{b}^{(\ell)} &:= \sigma^{-2} \bm{H}^{\rm T} \bm{y} + \bm{\gamma}^{(\ell)}. \nonumber
\end{align}
Solving a linear system is generally numerically more stable than computing the inverse, especially if the inverse is only part of a larger computation~\cite{croz1992stability}, as it is the case in~\eqref{eq:ep_mean_computation_with_inverse}.
Several stable and rigorously studied numerical methods exist to solve linear systems. We focus on iterative methods since they provide low-complexity approximative solutions upon early termination.
Besides the well-known \ac{GS} method~\cite[Sec.~11.2]{GvL13}, we consider Krylov subspace methods in the following. 

\subsubsection{Krylov Subspace Methods}
In numerical linear algebra, Krylov subspace methods are a well-understood tool to iteratively find approximations to solutions of linear systems. The key idea is to find the best approximation of the solution in a subspace, the so-called Krylov space. In each iteration, this subspace is enlarged and a better approximation is constructed in the larger Krylov subspace. Standard algorithms for the Krylov subspace method are the \ac{CG} algorithm and Lanczos/Arnoldi-based methods such as the \ac{FOM} and \ac{BiCG}~\cite{GvL13,SiSz2007}. The latter methods involve a solution of a ${k \times k}$ linear system in the $k^\text{th}$ step, which becomes inefficient for large~$k$. The \ac{CG} algorithm, which was originally published in \cite{hestenes1952cg}, is specifically tailored to symmetric positive definite problems like the matrices in~\eqref{eq:A_def}, and each iteration has the same computational complexity. However, the \ac{CG} algorithm is known to suffer from slow convergence if the problem is badly conditioned~\cite[Theorem 11.3.3]{GvL13}. We can substantially accelerate convergence by applying a preconditioner to the \ac{CG} that is specifically tailored to the characteristics of our considered problem. For more details on preconditioning, we refer the reader to~\cite[Sec.~11.5]{GvL13}.
\subsubsection{\Ac{pCG} Method}
Suppose we have a linear system ${\bm{Ax} = \bm{b}}$ for a symmetric, positive definite matrix ${\bm{A}\in \mathbb{R}^{N \times N}}$ and a matrix $\bm{B}$, such that ${\bm{B} \approx \bm{A}^{-1}}$. The matrix~$\bm{B}$ is called the preconditioner and has to be chosen adequately. For an arbitrary initialization $\bm{x}^{(0)}$, ${\bm{g}^{(0)} = \bm{A}\bm{x}^{(0)} - \bm{b}}$, ${\bm{d}^{(0)} = \bm{B} \bm{g}^{(0)}}$ and ${\rho^{(0)} = \langle \bm{B}\bm{g}^{(0)},\bm{g}^{(0)} \rangle}$ the \ac{pCG} algorithm reads for ${k=0,1,\dots}$ as 
\begin{align*}
    \bm{x}^{(k+1)} &= \bm{x}^{(k)} - \alpha^{(k)} \bm{d}^{(k)}, &&\bm{g}^{(k+1)} = \bm{g}^{(k)} - \alpha^{(k)} \bm{A} \bm{d}^{(k)}, \\
    \bm{d}^{(k+1)} &= \bm{B}\bm{g}^{(k+1)} + \beta^{(k)} \bm{d}^{(k)}, \hspace{-0.5em} &&\rho^{(k+1)} = \langle \bm{B}\bm{g}^{(k+1)},\bm{g}^{(k+1)} \rangle,
\end{align*}
where ${\alpha^{(k)} = {\rho^{(k)}}/{\langle \bm{A}\bm{d}^{(k)},\bm{d}^{(k)} \rangle}}$ and ${\beta^{(k)} = {\rho^{(k+1)}}/{\rho^{(k)}}}$. The computational cost per \ac{pCG} step is dominated by the matrix-vector product ${\bm{A}\bm{d}^{(k)}}$, which leads to a global complexity in the order of~${\mathcal{O}(kN^2)}$. The \ac{pCG} algorithm reaches (except for round-off errors) the exact solution of the linear system after ${k=N}$ iterations. For approximative solutions with ${k<N}$, a reasonable stopping criterion is ${\sqrt{\rho^{(k)}} \leq \text{tol} \cdot \| \bm{b}\|_2}$.

We apply the \ac{pCG} algorithm to solve the system in~\eqref{eq:linear_system}.
To obtain fast convergence of the \ac{pCG} method, the choice of a suitable preconditioner is crucial. In this special setting, the convergence of the \ac{EP} algorithm leads to decreasing marginal variances which implies large entries in $\bm{\Lambda}^{(\ell)}$ (cf.~\eqref{eq:lambda_update}) and hence strong diagonal dominance of $\bm{A}^{(\ell)}$. This behavior is also observed in our experiments. Hence, it is natural to choose a diagonal, so-called, Jacobi preconditioner \cite[Sec.~11.5]{GvL13}. For ${\bm{A}^{(\ell)}=(A_{nm})_{n,m=1}^{2N_\text{t}}}$, a good choice is $\bm{B}=\text{diag}(1/A_{nn})_{n=1}^{2N_t}$. This choice for $\bm{B}$ requires no matrix inversion and allows for very efficient matrix-vector products in the \ac{pCG} algorithm.

\subsection{Marginal Inference: Variance}\label{subsec:var_approx}
The marginal variances~$\sigma_n^{2(\ell)}$ are obtained in each iteration~$\ell$ of the \ac{EP} algorithm by extracting the diagonal elements of the covariance matrix~${\bm{\Sigma}^{(\ell)} = \bm{A}^{-1(\ell)}}$ in~\eqref{eq:matrix_inverse}.
To avoid the full matrix inversions for reduced complexity, we rely on an approximation similar to~\cite{tan_low-complexity_2019}, which is based on the Neumann-series expansion terminated after the first term:
\begin{equation}
    \sigma^{2(\ell)}_n = \Sigma^{(\ell)}_{nn} = \frac{1}{A^{(\ell)}_{nn}} + \sum\limits_{m \neq n} \frac{A^{(\ell)}_{nm} A^{(\ell)}_{mn}}{A^{(\ell)}_{mm}} \frac{1}{A^{2(\ell)}_{nn}} \approx \frac{1}{A^{(\ell)}_{nn}}. \label{eq:neumann_series}
\end{equation}
Inspired by~\cite{ge_improving_2021}, we introduce correction parameters to improve the approximation in~\eqref{eq:neumann_series}:
\begin{equation*}
    \tilde{\sigma}^{2(\ell)}_n := \max \left\{\alpha_1^{(\ell)} \frac{1}{A^{(\ell)}_{nn}} + \alpha_2^{(\ell)}, \frac{1}{A^{(\ell)}_{nn}} \right\}.
\end{equation*}
Compared to~\cite{ge_improving_2021}, the parameters apply a linear transformation on ${\frac{1}{A_{nn}}}$ (instead of on ${A_{nn}}$), and we lower bound $\tilde{\sigma}^2_n$ by ${\frac{1}{A_{nn}}}$ instead of by a constant (${\epsilon = 10^{-12}}$).
Optimizing $\alpha_1^{(\ell)}$ and $\alpha_2^{(\ell)}$ in each iteration~$\ell$ based on a labeled dataset ${\mathcal{D}^{(\ell)}_\text{train} = \{ (\bm{A}^{(\ell)}, \bm{A}^{-1(\ell)})_d, d=1,\ldots,D\}}$ of representative matrix-inverse pairs can be interpreted as linear regression\footnote{In numerical studies, we found that a piecewise exponential regression with two intervals can fit the data much better. However, the overall gains in terms of \ac{EP} detection performance were insignificant.} between the dependent variables~${(\bm{A}^{-1(\ell)})_{nn}=\sigma^{2(\ell)}_n}$ and the regressors~${\frac{1}{A^{(\ell)}_{nn}}}$.
Following the method of least squares, we obtain
\begin{equation*}
    \left( \alpha_1^{(\ell)}, \alpha_2^{(\ell)} \right) = \argmin_{(\alpha_1, \alpha_2) \in \mathbb{R}^2} \sum\limits_{\mathcal{D}^{(\ell)}_\text{train}} \sum\limits_{n=1}^{2N_\text{t}} \left( \alpha_1 \frac{1}{A^{(\ell)}_{nn}} + \alpha_2 - \sigma^{2(\ell)}_n \right)^2,
\end{equation*}
which can be minimized in closed form~\cite[Sec.~3]{bishop_pattern_2006}.
We observe that this approximation works remarkably well for \ac{EP} iterations ${\ell>0}$ and for ${\ell=0}$ if the Gram channel matrix~${\bm{H}^{\rm T}\bm{H}}$ is strongly diagonal dominant. In the remaining cases, the initialization with the inverse~$\bm{\Sigma}^{(0)}$ has shown to significantly increase accuracy.

\subsection{The Proposed EPiCG Algorithm}
Combining the discussions in Sec.~\ref{subsec:mean_approx} and~\ref{subsec:var_approx}, we propose the EPiCG algorithm  with two different initializations as our main contribution of this work.
We apply the original \ac{EP} detector (cf. Sec.~\ref{subsec:EPdetector}) with the following modifications:
\begin{itemize}
    \item EPiCG: Compute the means~${\mu_n^{(\ell)}}$ by solving the linear system in~\eqref{eq:linear_system} with the \ac{pCG} method and Jacobi-preconditioning. Approximate the variances~${\sigma_n^{2(\ell)}}$ by the strategy proposed in Sec.~\ref{subsec:var_approx}.
    \item EPiCG-$\bm{\Sigma}^{(0)}$: For ${\ell = 0}$, initialize ${\mu_n^{(0)}}$ and ${\sigma_n^{2(0)}}$ via the inverse~${\bm{A}^{-1(0)}=\bm{\Sigma}^{(0)}}$ as in the original \ac{EP} detector. Otherwise for ${\ell > 0}$, apply the EPiCG algorithm.
\end{itemize}

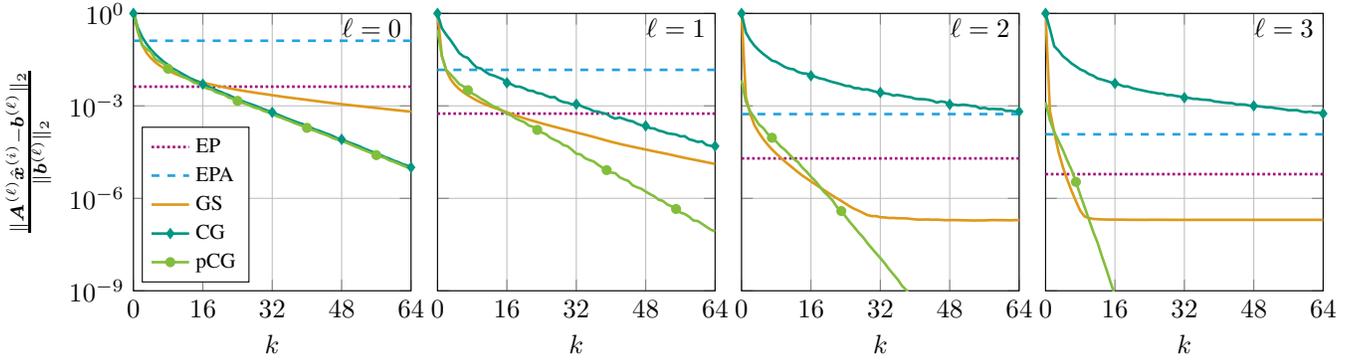
\begin{figure*}
    \centering
\begin{tikzpicture}
    \pgfplotsset{
legend image code/.code={
\draw[mark repeat=2,mark phase=2]
plot coordinates {
(0cm,0cm)
(0.25cm,0cm)        %
(0.5cm,0cm)         %
};%
}
}
\begin{groupplot}[group style={group size=4 by 1, horizontal sep=1em, vertical sep=1em},height=15em,width=15em, enlargelimits=false, grid=major, ymode=log, legend style={font=\footnotesize, cells={align=left}, anchor=south west, at={(0.03,0.03)}},legend cell align={left}]

\nextgroupplot[ymax=1.0, ymin=0.000000001, ytick={1.0, 0.001, 0.000001, 0.000000001}, xtick={0,16,32,48,64},
ylabel=$\frac{\lVert \bm{A}^{(\ell)} \hat{\bm{x}}^{(i)} - \bm{b}^{(\ell)}\rVert_2}{\lVert \bm{b}^{(\ell)} \rVert_2}$,
xlabel=$k$] %
    \addplot[color=EPcolor, densely dotted, line width=1pt] table[x={CG iters}, y={EPiter0 inverse median}, col sep=comma] {numerical_results/Rayleigh/approx_solvers/128x128_CGerror_over_iters_forall_EP_iters.csv};
    \addlegendentry{EP}
    \addplot[color=EPAcolor, dashed, line width=1pt] table[x={CG iters}, y={EPiter0 EPAapprox median}, col sep=comma] {numerical_results/Rayleigh/approx_solvers/128x128_CGerror_over_iters_forall_EP_iters.csv};
    \addlegendentry{EPA}
    \addplot[color=GScolor, line width=1pt] table[x={CG iters}, y={EPiter0 GS median}, col sep=comma] {numerical_results/Rayleigh/approx_solvers/128x128_CGerror_over_iters_forall_EP_iters.csv};
    \addlegendentry{GS}
    \addplot[color=CGcolor, line width=1pt, mark=diamond*, mark options={scale=0.7, solid}, mark repeat=16, mark phase=1] table[x={CG iters}, y={EPiter0 CG median}, col sep=comma] {numerical_results/Rayleigh/approx_solvers/128x128_CGerror_over_iters_forall_EP_iters.csv};
    \addlegendentry{CG}
    \addplot[color=pCGcolor, line width=1pt, mark=*, mark options={scale=0.7, solid}, mark repeat=16, mark phase=8] table[x={CG iters}, y={EPiter0 pCG median}, col sep=comma] {numerical_results/Rayleigh/approx_solvers/128x128_CGerror_over_iters_forall_EP_iters.csv};
    \addlegendentry{pCG}
    \node[coordinate] (A) at (64,0.35) {};
    \node[left=0em of A.center, anchor=east](B){${\ell=0}$};
    \node[coordinate] (C) at (64,0.04) {};
\nextgroupplot[yticklabels={}, ymax=1.0, ymin=0.000000001, ytick={1.0, 0.001, 0.000001, 0.000000001}, xtick={0,16,32,48,64},
xlabel=$k$] %
    \addplot[color=EPcolor, densely dotted, line width=1pt] table[x={CG iters}, y={EPiter1 inverse median}, col sep=comma] {numerical_results/Rayleigh/approx_solvers/128x128_CGerror_over_iters_forall_EP_iters.csv};
    \addplot[color=EPAcolor, dashed, line width=1pt] table[x={CG iters}, y={EPiter1 EPAapprox median}, col sep=comma] {numerical_results/Rayleigh/approx_solvers/128x128_CGerror_over_iters_forall_EP_iters.csv};
    \addplot[color=GScolor, line width=1pt] table[x={CG iters}, y={EPiter1 GS median}, col sep=comma] {numerical_results/Rayleigh/approx_solvers/128x128_CGerror_over_iters_forall_EP_iters.csv};
    \addplot[color=CGcolor, line width=1pt, mark=diamond*, mark options={scale=0.7, solid}, mark repeat=16, mark phase=1] table[x={CG iters}, y={EPiter1 CG median}, col sep=comma] {numerical_results/Rayleigh/approx_solvers/128x128_CGerror_over_iters_forall_EP_iters.csv};
    \addplot[color=pCGcolor, line width=1pt, mark=*, mark options={scale=0.7, solid}, mark repeat=16, mark phase=8] table[x={CG iters}, y={EPiter1 pCG median}, col sep=comma] {numerical_results/Rayleigh/approx_solvers/128x128_CGerror_over_iters_forall_EP_iters.csv};
    \node[coordinate] (A) at (64,0.35) {};
    \node[left=0em of A.center, anchor=east](B){${\ell=1}$};
    \node[coordinate] (C) at (64,0.04) {};

\nextgroupplot[yticklabels={}, ymax=1.0, ymin=0.000000001, ytick={1.0, 0.001, 0.000001, 0.000000001}, xtick={0,16,32,48,64},
xlabel=$k$] %
    \addplot[color=EPcolor, densely dotted, line width=1pt] table[x={CG iters}, y={EPiter2 inverse median}, col sep=comma] {numerical_results/Rayleigh/approx_solvers/128x128_CGerror_over_iters_forall_EP_iters.csv};
    \addplot[color=EPAcolor, dashed, line width=1pt] table[x={CG iters}, y={EPiter2 EPAapprox median}, col sep=comma] {numerical_results/Rayleigh/approx_solvers/128x128_CGerror_over_iters_forall_EP_iters.csv};
    \addplot[color=GScolor, line width=1pt] table[x={CG iters}, y={EPiter2 GS median}, col sep=comma] {numerical_results/Rayleigh/approx_solvers/128x128_CGerror_over_iters_forall_EP_iters.csv};
    \addplot[color=CGcolor, line width=1pt, mark=diamond*, mark options={scale=0.7, solid}, mark repeat=16, mark phase=1] table[x={CG iters}, y={EPiter2 CG median}, col sep=comma] {numerical_results/Rayleigh/approx_solvers/128x128_CGerror_over_iters_forall_EP_iters.csv};
    \addplot[color=pCGcolor, line width=1pt, mark=*, mark options={scale=0.7, solid}, mark repeat=16, mark phase=8] table[x={CG iters}, y={EPiter2 pCG median}, col sep=comma] {numerical_results/Rayleigh/approx_solvers/128x128_CGerror_over_iters_forall_EP_iters.csv};
    \node[coordinate] (A) at (64,0.35) {};
    \node[left=0em of A.center, anchor=east](B){${\ell=2}$};
    \node[coordinate] (C) at (64,0.04) {};

\nextgroupplot[yticklabels={}, ymax=1.0, ymin=0.000000001, ytick={1.0, 0.001, 0.000001, 0.000000001}, xtick={0,16,32,48,64},
xlabel=$k$, legend style={font=\footnotesize, cells={align=left}, anchor=south, at={(0.5,0.02)}}] %
    \addplot[color=EPcolor, densely dotted, line width=1pt] table[x={CG iters}, y={EPiter3 inverse median}, col sep=comma] {numerical_results/Rayleigh/approx_solvers/128x128_CGerror_over_iters_forall_EP_iters.csv};
    \addplot[color=EPAcolor, dashed, line width=1pt] table[x={CG iters}, y={EPiter3 EPAapprox median}, col sep=comma] {numerical_results/Rayleigh/approx_solvers/128x128_CGerror_over_iters_forall_EP_iters.csv};
    \addplot[color=GScolor, line width=1pt] table[x={CG iters}, y={EPiter3 GS median}, col sep=comma] {numerical_results/Rayleigh/approx_solvers/128x128_CGerror_over_iters_forall_EP_iters.csv};
    \addplot[color=CGcolor, line width=1pt, mark=diamond*, mark options={scale=0.7, solid}, mark repeat=16, mark phase=1] table[x={CG iters}, y={EPiter3 CG median}, col sep=comma] {numerical_results/Rayleigh/approx_solvers/128x128_CGerror_over_iters_forall_EP_iters.csv};
    \addplot[color=pCGcolor, line width=1pt, mark=*, mark options={scale=0.7, solid}, mark repeat=16, mark phase=8] table[x={CG iters}, y={EPiter3 pCG median}, col sep=comma] {numerical_results/Rayleigh/approx_solvers/128x128_CGerror_over_iters_forall_EP_iters.csv};
    \node[coordinate] (A) at (64,0.35) {};
    \node[left=0em of A.center, anchor=east](B){${\ell=3}$};
    \node[coordinate] (C) at (64,0.04) {};

\end{groupplot}
\end{tikzpicture}
    \caption{Median RRE of various approximative solvers over the steps~$k$. The plots vary the \ac{EP} iteration~$\ell$, in which the linear system~\eqref{eq:linear_system} is solved.
    }
    \label{fig:CGiters}
\end{figure*}
\section{Experimental Study}\label{sec:results}
We evaluate the EPiCG algorithm for \ac{MIMO} detection and focus in particular on the analysis and distinction between the two proposed approximations. In the following numerical experiments, we use a $16$-\ac{QAM} constellation and fix the number of receive antennas to~${N_\text{r}=128}$. To cover different \ac{MUI} scenarios, we study both ${N_\text{t}=64}$ and ${N_\text{t}=128}$ single-antenna users. The results are based on $5000$ channel matrices~$\cs{\bm{H}}$ that were randomly sampled from the channel models introduced in Sec.~\ref{subsec:channel_model}, respectively.

\subsection{Approximation of the Marginal Mean} %
We start by evaluating the suitability of different approximative methods for solving the linear system~\eqref{eq:linear_system} in each \ac{EP} iteration~$\ell$. Figure~\ref{fig:CGiters} plots the \ac{RRE}~$\frac{\lVert \bm{A}^{(\ell)} \hat{\bm{x}}^{(k)} - \bm{b}^{(\ell)}\rVert_2}{\lVert \bm{b}^{(\ell)} \rVert_2}$ over the steps~$k$ of the \ac{GS}, \ac{CG} and \ac{pCG} method, respectively.
The results are shown for the Rayleigh fading channel with ${N_\text{t}=128}$ and ${\mathsf{snr}=19\,}$dB. %
We compare the methods with two non-iterative approaches from literature: solving the linear system using the inverse ${\hat{\bm{x}} = \bm{A}^{-1(\ell)}\bm{b}^{(\ell)}}$ as proposed in the original \ac{EP} detector~\cite{cespedes_expectation_2014}, and the approximation used in the EPA algorithm, i.e., avoiding to solve the linear system completely by assuming ${\hat{x}_n = \mu_{p_n}^{(\ell)} }$. However, this assumption is not correct, but only applies when the fixed points of \ac{EP} are reached~\cite{tan_low-complexity_2019}.

In \ac{EP} iteration~${\ell=0}$, the \ac{GS} solver is outperformed by the \ac{CG} and \ac{pCG} method for~${k>16}$. The latter behave almost identically. The benefit of the Jacobi preconditioner becomes apparent for the subsequent \ac{EP} iterations ${\ell>0}$, where the matrix $\bm{A}^{(\ell)}$ becomes more and more diagonal-dominant. 
While the conventional \ac{CG} solver converges similarly or slower compared to~${\ell=0}$, the \ac{pCG} method significantly increases the convergence rate in each subsequent \ac{EP} iteration~$\ell$. 
Since the computational complexity scales quadratically for each step of the considered solvers, the plots indicate a precision-complexity tradeoff.
For instance in~${\ell=1}$, the \ac{pCG} solver only requires ${k=16}$ steps to reach the same $\text{\ac{RRE}}=10^{-4}$ as the inverse-based solution of the \ac{EP} detector. 
Note that the \ac{RRE} is inherently available within the \ac{pCG} steps which offers great flexibility for controlling this tradeoff.

The approximation of the EPA algorithm is very coarse in ${\ell=0}$ but quickly improves with the convergence of the \ac{EP} algorithm, i.e., with increasing~$\ell$. It approximatively requires ${k=4}$ iterations of the \ac{pCG} algorithm to reach the \ac{RRE} of the EPA algorithm in each \ac{EP} iteration~$\ell$.

\subsection{Detection Performance}
We compare the performance of the proposed EPiCG algorithm to the \ac{LMMSE} detector, the EPANet algorithm~\cite{ge_improving_2021} as well as to the original \ac{EP} detector~\cite{cespedes_expectation_2014}. We fix ${\beta=0.1}$ for the \ac{EP} and EPiCG algorithm and we perform ${L=10}$ iterations for all iterative detection schemes. We also consider the initialization variant of the EPiCG-$\bm{\Sigma}^{(0)}$ algorithm for the EPANet algorithm as suggested in~\cite{ge_improving_2021}, accordingly denoted by~{${\text{EPANet-}\bm{\Sigma}^{(0)}}$.
We fix the stopping criterion of the \ac{pCG} algorithm to $\text{tol}(\ell)=(10^{-3},10^{-4},10^{-5},10^{-5},\ldots,10^{-5})$. For the Rayleigh-fading channel, the precision is thereby similar to the matrix inverse (cf. Fig.~\ref{fig:CGiters}) which enables a good comparison since the main difference between the EPiCG algorithm and the \ac{EP} detector stems from the variance approximation.
To create $\mathcal{D}_\text{train}$, we randomly generate ${D=100}$ transmission scenarios with $\mathsf{snr}$ values sampled from a uniform distribution with bounds chosen such that the \ac{SER} of the resulting EPiCG detector performance is in the target \ac{SER} interval $\left[10^{-2};10^{-1}\right]$.

\begin{figure}
    \centering
    \pgfplotsset{
compat=1.11,
legend image code/.code={
\draw[mark repeat=2,mark phase=2]
plot coordinates {
(0cm,0cm)
(0.25cm,0cm)        %
(0.5cm,0cm)         %
};%
}
}
    \begin{tikzpicture}
    \begin{axis}[
    width=\linewidth, %
    height=0.75\linewidth,
    align = left,
    grid=major, %
    grid style={gray!30}, %
    xlabel= $\mathsf{snr}$ (dB),
    ylabel= SER,
	  scaled y ticks=false,
    ymode = log,
    ymin = 0.001,
    ymax = 1.0,
    xmin = 6,
    xmax = 22,
    enlarge x limits=false,
    enlarge y limits=false,
    line width=1pt,
	  legend style={font=\footnotesize, cells={align=left}, anchor=south west, at={(0.018,0.018)}},
    legend cell align={left},
	  smooth,
    ]
    \addplot[color=gray, dashdotted, line width=1pt] table[x={SNR dB}, y={LMMSE SER}, col sep=comma] {numerical_results/Rayleigh/SERvsSNR/128x128.csv};
    \addlegendentry{LMMSE}
    \addplot[draw=none,color=EPcolor, densely dotted, line width=1pt] table[x={SNR dB}, y={EP SER}, col sep=comma] {numerical_results/Rayleigh/SERvsSNR/128x128.csv};
    \addlegendentry{EP}
    \addplot[color=EPAcolor, line width=1pt, mark=square*, mark options={scale=0.7, solid}, mark repeat=2] table[x={SNR dB}, y={EPANet SER}, col sep=comma] {numerical_results/Rayleigh/SERvsSNR/128x128.csv};
    \addlegendentry{EPANet}
    \addplot[color=EPAcolor, dashed, line width=1pt, mark=square*, mark options={scale=0.7, solid}, mark repeat=2, mark phase=2] table[x={SNR dB}, y={EPANet (LMMSE init) SER}, col sep=comma] {numerical_results/Rayleigh/SERvsSNR/128x128.csv};
    \addlegendentry{EPANet-$\bm{\Sigma}^{(0)}$}
    \addplot[color=EPiCGcolor, line width=1pt, mark=*, mark options={scale=0.7, solid}, mark repeat=2] table[x={SNR dB}, y={EPiCG SER}, col sep=comma] {numerical_results/Rayleigh/SERvsSNR/128x128.csv};
    \addlegendentry{EPiCG}
    \addplot[color=EPiCGcolor, dashed, line width=1pt, mark=*, mark options={scale=0.7, solid}, mark repeat=2, mark phase=2] table[x={SNR dB}, y={EPiCG (LMMSE init) SER}, col sep=comma] {numerical_results/Rayleigh/SERvsSNR/128x128.csv};
    \addlegendentry{EPiCG-$\bm{\Sigma}^{(0)}$}
    \addplot[color=EPcolor, densely dotted, line width=1pt] table[x={SNR dB}, y={EP SER}, col sep=comma] {numerical_results/Rayleigh/SERvsSNR/128x128.csv};

    \addplot[color=gray, dashdotted, line width=1pt] table[x={SNR dB}, y={LMMSE SER}, col sep=comma] {numerical_results/Rayleigh/SERvsSNR/64x128.csv};

    \addplot[color=EPAcolor, line width=1pt, mark=square*, mark options={scale=0.7, solid}, mark repeat=2, mark phase=1] table[x={SNR dB}, y={EPANet SER}, col sep=comma] {numerical_results/Rayleigh/SERvsSNR/64x128.csv};

    \addplot[color=EPiCGcolor, line width=1pt, mark=*, mark options={scale=0.7, solid}, mark repeat=2, mark phase=2] table[x={SNR dB}, y={EPiCG SER}, col sep=comma] {numerical_results/Rayleigh/SERvsSNR/64x128.csv};

    \addplot[color=EPcolor, densely dotted, line width=1pt] table[x={SNR dB}, y={EP SER}, col sep=comma] {numerical_results/Rayleigh/SERvsSNR/64x128.csv};

    \node[coordinate] (A) at (8.9,0.26) {};
    \draw [line width=0.5pt] (A) ellipse [x radius=0.8em, y radius=0.3em, rotate=45];
    \node[below left=0.7em and 0.7em of A.south, anchor=north](32x64){${\scriptstyle {N_\text{t}=64}}$};
    \node[coordinate] (B) at (13.87,0.42) {};
    \draw [line width=0.5pt] (B) ellipse [x radius=0.8em, y radius=0.3em, rotate=45];
    \node[below left=0.7em and 0.7em of B.south, anchor=north](32x64){${\scriptstyle {N_\text{t}=128}}$};
  \end{axis}
\end{tikzpicture}
    \caption{\ac{SER} over $\mathsf{snr}$ for various \ac{MIMO} detectors on the Rayleigh-fading channel with ${N_\text{t}=64,128}$ and ${N_\text{r}=128}$.}
    \label{fig:BER_Rayleigh_Nr128}
\end{figure}
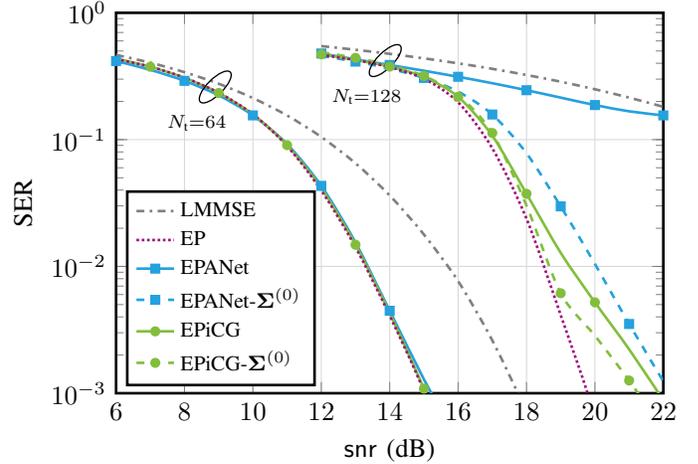
Figure~\ref{fig:BER_Rayleigh_Nr128} evaluates the \ac{SER} over the \ac{SNR} for the Rayleigh-fading channel. In the low-\ac{MUI} scenario with ${N_\text{t}=64}$, both the EPANet and the EPiCG algorithm approach the \ac{EP} detector performance. %
In the high-\ac{MUI} scenario with ${N_\text{t}=N_\text{r}}$, the EPANet detector behaves similarly to the \ac{LMMSE} detector and only reaches competitive performance when being initialized with~$\bm{\Sigma}^{(0)}$. The EPiCG detector follows the waterfall behavior of the \ac{EP} detector with an $\mathsf{snr}$ offset of $0.75\,$dB for ${\text{SER}=10^{-2}}$. We observed that the remaining gap between the EPiCG algorithm compared to the \ac{EP} detector comes from the variance approximation in iteration~${\ell=0}$ as the dominant source of error. The EPiCG-$\bm{\Sigma}^{(0)}$ algorithm consequently closes this gap.

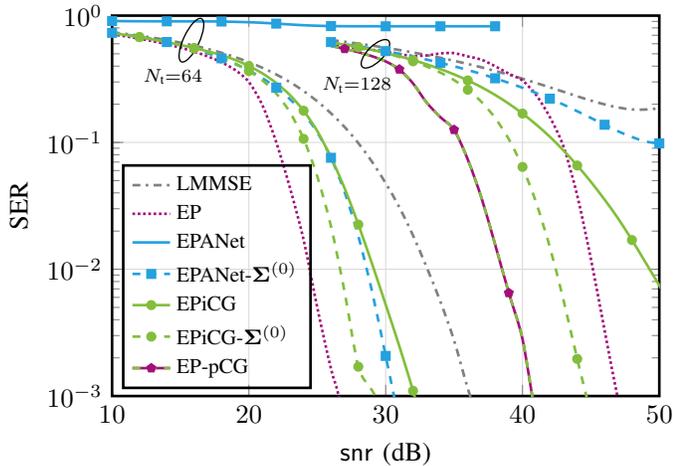
\begin{figure}
    \centering
    \pgfplotsset{
compat=1.11,
legend image code/.code={
\draw[mark repeat=2,mark phase=2]
plot coordinates {
(0cm,0cm)
(0.25cm,0cm)        %
(0.5cm,0cm)         %
};%
}
}
    \begin{tikzpicture}
    \begin{axis}[
    width=\linewidth, %
    height=0.75\linewidth,
    align = left,
    grid=major, %
    grid style={gray!30}, %
    xlabel= $\mathsf{snr}$ (dB),
    ylabel= SER,
	  scaled y ticks=false,
    ymode = log,
    ymin = 0.001,
    ymax = 1.0,
    xmin = 10,
    xmax = 50,
    enlarge x limits=false,
    enlarge y limits=false,
    line width=1pt,
	  legend style={font=\footnotesize, cells={align=left}, anchor=south west, at={(0.018,0.018)}, fill=none},
    legend cell align={left},
	  smooth,
    ]
    \addplot[color=gray, dashdotted, line width=1pt] table[x={SNR dB}, y={LMMSE SER}, col sep=comma] {numerical_results/MIMO3D_UMa_NLOS/SERvsSNR/128x128.csv};
    \addlegendentry{LMMSE}
    \addplot[color=EPcolor, densely dotted, line width=1pt] table[x={SNR dB}, y={EP SER}, col sep=comma] {numerical_results/MIMO3D_UMa_NLOS/SERvsSNR/128x128.csv};
    \addlegendentry{EP}
    \addplot[draw=none,color=EPAcolor, line width=1pt] table[x={SNR dB}, y={EPANet SER}, col sep=comma]{numerical_results/MIMO3D_UMa_NLOS/SERvsSNR/128x128.csv};
    \addlegendentry{EPANet}
    \addplot[color=EPAcolor, dashed, line width=1pt, mark=square*, mark options={scale=0.7, solid}, mark repeat=4, mark phase=1] table[x={SNR dB}, y={EPANet (LMMSE init) SER}, col sep=comma] {numerical_results/MIMO3D_UMa_NLOS/SERvsSNR/128x128.csv};
    \addlegendentry{EPANet-$\bm{\Sigma}^{(0)}$}
    \addplot[color=EPiCGcolor, line width=1pt, mark=*, mark options={scale=0.7, solid}, mark repeat=4, mark phase=3] table[x={SNR dB}, y={EPiCG SER}, col sep=comma] {numerical_results/MIMO3D_UMa_NLOS/SERvsSNR/128x128.csv};
    \addlegendentry{EPiCG}
    \addplot[color=EPiCGcolor, dashed, line width=1pt, mark=*, mark options={scale=0.7, solid}, mark repeat=4, mark phase=3] table[x={SNR dB}, y={EPiCG (LMMSE init) SER}, col sep=comma] {numerical_results/MIMO3D_UMa_NLOS/SERvsSNR/128x128.csv};
    \addlegendentry{EPiCG-$\bm{\Sigma}^{(0)}$}
    \addplot[color=EPcolor, line width=1pt, postaction={dashed, color=EPiCGcolor}, mark=pentagon*, mark options={scale=0.7, solid}, mark repeat=4, mark phase=2] table[x={SNR dB}, y={EPsolve SER}, col sep=comma] {numerical_results/MIMO3D_UMa_NLOS/SERvsSNR/128x128.csv};
    \addlegendentry{EP-pCG}

    \addplot[color=gray, dashdotted, line width=1pt] table[x={SNR dB}, y={LMMSE SER}, col sep=comma] {numerical_results/MIMO3D_UMa_NLOS/SERvsSNR/64x128.csv};
    \addplot[color=EPcolor, densely dotted, line width=1pt] table[x={SNR dB}, y={EP SER}, col sep=comma] {numerical_results/MIMO3D_UMa_NLOS/SERvsSNR/64x128.csv};
    \addplot[color=EPAcolor, line width=1pt, mark=square*, mark options={scale=0.7, solid}, mark repeat=4, mark phase=1] table[x={SNR dB}, y={EPANet SER}, col sep=comma]{numerical_results/MIMO3D_UMa_NLOS/SERvsSNR/64x128.csv};
    \addplot[color=EPAcolor, dashed, line width=1pt, mark=square*, mark options={scale=0.7, solid}, mark repeat=4, mark phase=1] table[x={SNR dB}, y={EPANet (LMMSE init) SER}, col sep=comma] {numerical_results/MIMO3D_UMa_NLOS/SERvsSNR/64x128.csv};
    \addplot[color=EPiCGcolor, line width=1pt, mark=*, mark options={scale=0.7, solid}, mark repeat=4, mark phase=3] table[x={SNR dB}, y={EPiCG SER}, col sep=comma] {numerical_results/MIMO3D_UMa_NLOS/SERvsSNR/64x128.csv};
    \addplot[color=EPiCGcolor, dashed, line width=1pt, mark=*, mark options={scale=0.7, solid}, mark repeat=4, mark phase=3] table[x={SNR dB}, y={EPiCG (LMMSE init) SER}, col sep=comma] {numerical_results/MIMO3D_UMa_NLOS/SERvsSNR/64x128.csv};

    \node[coordinate] (A) at (15.8,0.64) {};
    \draw [line width=0.5pt] (A) ellipse [x radius=0.9em, y radius=0.3em, rotate=65];
    \node[below left=0.7em and 0.7em of A.south, anchor=north](32x64){${\scriptstyle {N_\text{t}=64}}$};
    \node[coordinate] (B) at (29.25,0.5) {};
    \draw [line width=0.5pt] (B) ellipse [x radius=0.7em, y radius=0.3em, rotate=45];
    \node[below left=0.5em and 0.7em of B.south, anchor=north](32x64){${\scriptstyle {N_\text{t}=128}}$};
  \end{axis}
\end{tikzpicture}
    \caption{\ac{SER} over $\mathsf{snr}$ for various \ac{MIMO} detectors on the 3GPP 3D \ac{MIMO} UMa \ac{NLOS} channel with ${N_\text{t}=64,128}$ and ${N_\text{r}=128}$.}
    \label{fig:BER_UMa}
\end{figure}

We further evaluate the performance of the considered detectors on the 3GPP 3D MIMO \ac{UMa} \ac{NLOS} channel model in Fig.~\ref{fig:BER_UMa}. The significantly increased correlation in this channel model and the consequently decreased diagonal dominance of the Gram channel matrices generally degrades the performance of both the EPANet detector and the EPiCG algorithm. For ${N_\text{t}=64}$, the EPiCG-$\bm{\Sigma}^{(0)}$ algorithm reduces the $\mathsf{snr}$ gap to \ac{EP} detection from $4\,$dB to $2\,$dB at an ${\text{SER}=10^{-2}}$, compared to the EPiCG and EPANet-$\bm{\Sigma}^{(0)}$ detectors.
In the high-\ac{MUI} scenario with ${N_\text{t}=128}$, the EPiCG-$\bm{\Sigma}^{(0)}$ algorithm outperforms the \ac{EP} detector. We conjecture that the performance of the \ac{EP} detector degrades due to an unstable solution of eq.~\eqref{eq:ep_mean_computation_with_inverse} via the matrix inverse. To verify this, we replace the computation of the mean in the \ac{EP} detector by the \ac{pCG} algorithm with ${\text{tol}=10^{-5}}$ and only use the diagonal elements of the matrix inverse to obtain the variances. The result is denoted by EP-pCG in Fig.~\ref{fig:BER_UMa} and confirms the conjecture.

\subsection{Complexity Analysis}
Finally, we evaluate the proposed EPiCG detection algorithms from a computational complexity perspective. Table~\ref{tab:complexity} provides the total number of \ac{pCG} steps that are required within the EPiCG algorithm over all \ac{EP} iterations, i.e., to solve $L$ linear systems of dimension ${2N_\text{t}\times 2N_\text{t}}$. Since we use a precision-based stopping criterion instead of a fixed number of \ac{pCG} steps, the numbers represent rounded mean values over all $5000$ evaluation samples.
The EPiCG-$\bm{\Sigma}^{(0)}$ variant uses the matrix inverse for the computation of the marginal means in \ac{EP} iteration~${\ell=0}$ and requires approximately half of the \ac{pCG} steps compared with the EPiCG algorithm.
Note that optimizing the distribution of the \ac{pCG} steps over the \ac{EP} iterations might have the potential to improve further the performance-complexity tradeoff, which we leave for future work.

\section{Conclusion}
We propose the EpiCG algorithm as a faster and more robust method for massive \ac{MIMO} detection.
Replacing the matrix inversion for marginal inference in the \ac{EP} algorithm by the application of the \ac{pCG} method reduces the computational complexity from $\mathcal{O}(N^3)$ to $\mathcal{O}(kN^2)$ operations in each \ac{EP} iteration.
This offers a flexible performance-complexity tradeoff, while the accuracy is only mildly reduced due to an approximation of the marginal variance. 
Simulations on realistic \ac{MIMO} channels illustrate that the \ac{pCG} algorithm furthermore stabilizes the computation compared to the matrix inversion for highly-correlated channels with strong \ac{MUI}.
\begin{table}[t] 
\begin{center}
        \caption{\sc \centering Total number of \ac{pCG} steps in all ${L=10}$ \ac{EP} iterations of the EPiCG algorithm for different transmission scenarios.}
    \begin{tabular}{l rr rr rr}
    \toprule
        \multicolumn{1}{l}{Algorithm} & \multicolumn{2}{c}{EPiCG} & \multicolumn{1}{c}{}  & \multicolumn{2}{c}{EPiCG-$\bm{\Sigma}^{(0)}$} \\
         \multicolumn{1}{l}{${N_\text{t}}$} & $64$ & $128$ & & $64$ & $128$ \\
         \midrule
         Rayleigh & 40 & 125 & & 28 & 94  \\
         3GPP \ac{UMa} \ac{NLOS} & 187 & 652 & & 98 & 323 \\
         \bottomrule
    \end{tabular}
    \label{tab:complexity}
\end{center}
\end{table}

\end{document}